\documentclass[12pt, a4paper]{article}

\makeatletter
\renewcommand\footnotesize{%
   \@setfontsize\footnotesize\@xipt\@xipt
   \abovedisplayskip 10\p@ \@plus2\p@ \@minus5\p@
   \abovedisplayshortskip \z@ \@plus3\p@
   \belowdisplayshortskip 6\p@ \@plus3\p@ \@minus3\p@
   \def\@listi{marginmargini
               \topsep 6\p@ \@plus2\p@ \@minus2\p@
               \parsep 3\p@ \@plus2\p@ \@minus\p@
               \itemsep \parsep}%
   \belowdisplayskip \abovedisplayskip
}
\makeatother

\usepackage{comment}

\usepackage[american]{babel}

\usepackage[babel]{csquotes}

\usepackage[authordate, natbib, isbn=false, backend=biber, noibid]{biblatex-chicago}  
\usepackage{authblk}

\usepackage[bottom]{footmisc}

\usepackage{amsmath}
\usepackage{amsfonts}

\usepackage{caption}
\usepackage{geometry} 
\usepackage{setspace}
\usepackage{etoolbox}
\newcommand{\zerodisplayskips}{%
  \setlength{\abovedisplayskip}{0pt}%
  \setlength{\belowdisplayskip}{12pt}%
  \setlength{\abovedisplayshortskip}{0pt}%
  \setlength{\belowdisplayshortskip}{12pt}}
\appto{\normalsize}{\zerodisplayskips}
\appto{\small}{\zerodisplayskips}
\appto{\footnotesize}{\zerodisplayskips}

\usepackage{graphicx}

\usepackage{booktabs} 

\usepackage{array}

\usepackage{paralist} 

\usepackage{verbatim}

\usepackage{subfig} 

\usepackage{fancyhdr} 
\usepackage[nottoc, notlof, notlot]{tocbibind} 
\usepackage[titles, subfigure]{tocloft} 

\begin{document}

\thispagestyle{empty}

\bigskip

\begin{center}
{\Large Presenting the Probabilities of Different Effect Sizes: Towards a Better Understanding and Communication of Statistical Uncertainty}
\end{center}

\bigskip

\begin{center}
{\large Akisato Suzuki}

\begin{singlespace}
School of Politics and International Relations\\
University College Dublin\\
Belfield, Dublin 4, Ireland\\
akisato.suzuki@gmail.com\\
ORCID: 0000-0003-3691-0236
\end{singlespace}
\end{center}

\bigskip

\begin{center}
{\large Working paper\\ (November 8 2022)}
\end{center}

\begin{singlespace}

\begin{center}
\textbf{Abstract}
\end{center}

\noindent
How should social scientists understand and communicate the uncertainty of statistically estimated causal effects? I propose we utilize the posterior distribution of a causal effect and present the probability of the effect being greater (in absolute terms) than different minimum effect sizes. Probability is an intuitive measure of uncertainty for understanding and communication. In addition, the proposed approach needs no decision threshold for an uncertainty measure or an effect size, unlike the conventional approaches. I apply the proposed approach to a previous social scientific study, showing it enables richer inference than the significance-vs.-insignificance approach taken by the original study. The accompanying R package makes my approach easy to implement.
\end{singlespace}


\newpage

\section{Introduction}
How should social scientists understand and communicate the uncertainty of statistically estimated causal effects? It is common to use a specific decision threshold of a \textit{p}-value or confidence/credible interval (usually a p-value of 5\% and a 95\% interval) to conclude whether a causal effect is significant or not. While convenient to make categorical decisions, the significance-vs.-insignificance dichotomy leads us to overlook the full nuance of the statistical measure of uncertainty. This is because uncertainty is the degree of confidence and, therefore, a continuous scale.

I propose we utilize the posterior distribution of a causal effect, and present the probability of the effect being greater than different minimum effect sizes, by the plot of a complementary cumulative distribution (here, ``greater'' is meant in absolute terms: a greater positive value than zero or some positive value, or a greater negative value than zero or some negative value). In this way, it is unnecessary for researchers to use any decision threshold for the ``significance,'' ``confidence,'' or ``credible'' level of uncertainty, or for the effect size beyond which the effect is considered practically relevant \autocite{Gross2015, Kruschke2018a}. Researchers play the role of information providers and present different effect sizes and their associated probabilities as such. The approach is applicable regardless of the types of causal effect estimate (population average, sample average, individual, etc.).

The proposed approach should be used and interpreted, with a holistic view including the rigor of a causal theory and research design. For example, if a study used a convincing causal mechanism and research design, but if it had access to a small sample only and produced a degree of uncertainty greater than the conventional threshold for statistical significance, it could still make a case for attention. Put differently, the proposed approach should not be used to allow theoretically or methodologically questionable research to claim credibility based on not so high a probability \autocite[see][]{Lakens2016}. In short, the premise of the proposed approach is that statistically estimated causal effects are not based on questionable research. In addition, the proposed approach may not be the most effective way to present uncertainty in all circumstances. Further research is necessary to understand what the best practice is to present and interpret statistical uncertainty under what contexts.

In the rest of the article, I first elaborate on the proposed approach. Then, as an example, I apply the approach to \textcite{Huff2016}, an experimental study on the effect of protest methods on public opinion. The accompanying R package makes the proposed approach easy to implement (see the section ``Supplemental Materials''). All statistical analyses for this article were done on RStudio \autocite{RStudioTeam2020} running R version 4.1.2 \autocite{RCoreTeam2021}. The data visualization was done by the ggplot2 package \autocite{Wickham2016}.

\section{The Proposed Approach}
My proposed approach utilizes a posterior distribution estimated by Bayesian statistics \autocite[for Bayesian statistics, see, for example,][]{Gelman2013BDA, Gill2015, Kruschke2015, McElreath2016}. A posterior, denoted as $p(\theta|D,\ M)$, is the probability distribution of a parameter, $\theta$, given data, $D$, and model assumptions, $M$, such as a functional form, an identification strategy, and a prior distribution of $\theta$. Here, let us assume $\theta$ is a parameter for the effect of a causal factor.

I apply a complementary cumulative distribution function to $\theta$. More specifically, it computes the probability that a causal factor has an effect greater (in absolute terms) than some effect size. Formally: $P(\theta>\tilde\theta^{+}| D, M)$ for the positive values of $\theta$, where $\tilde\theta^{+}$ is zero or some positive value; $P(\theta<\tilde\theta^{-}| D, M)$ for the negative values of $\theta$, where $\tilde\theta^{-}$ is zero or some negative value. If we compute this probability while changing $\tilde\theta$ up to the theoretical or practical limits (e.g., up to the theoretical bounds of a posterior or up to the minimum/maximum in posterior samples), we obtain a complementary cumulative distribution. Experimental studies suggest that the plot of a complementary cumulative distribution enables relatively accurate interpretation of uncertainty information and effective decision making under uncertainty \autocites{Allen2014, Edwards2012, Fernandes2018}.

A posterior distribution might include both positive and negative values. In such a case, we should compute two complementary cumulative distribution functions, one for the positive values of $\theta$, i.e., $P(\theta>\tilde\theta^{+}| D, M)$ and the other for the negative values of $\theta$, i.e., $P(\theta<\tilde\theta^{-}| D, M)$. Whether it is plausible to suspect the effect can be either positive or negative, depends on the causal theory being drawn from. If only either direction of the effect is theoretically plausible \autocite[e.g., see][]{Vanderweele2010}, this should be reflected on the prior of $\theta$, e.g., by putting a bound on the range of the parameter.

\begin{figure}[t]
  \includegraphics[scale=0.15]{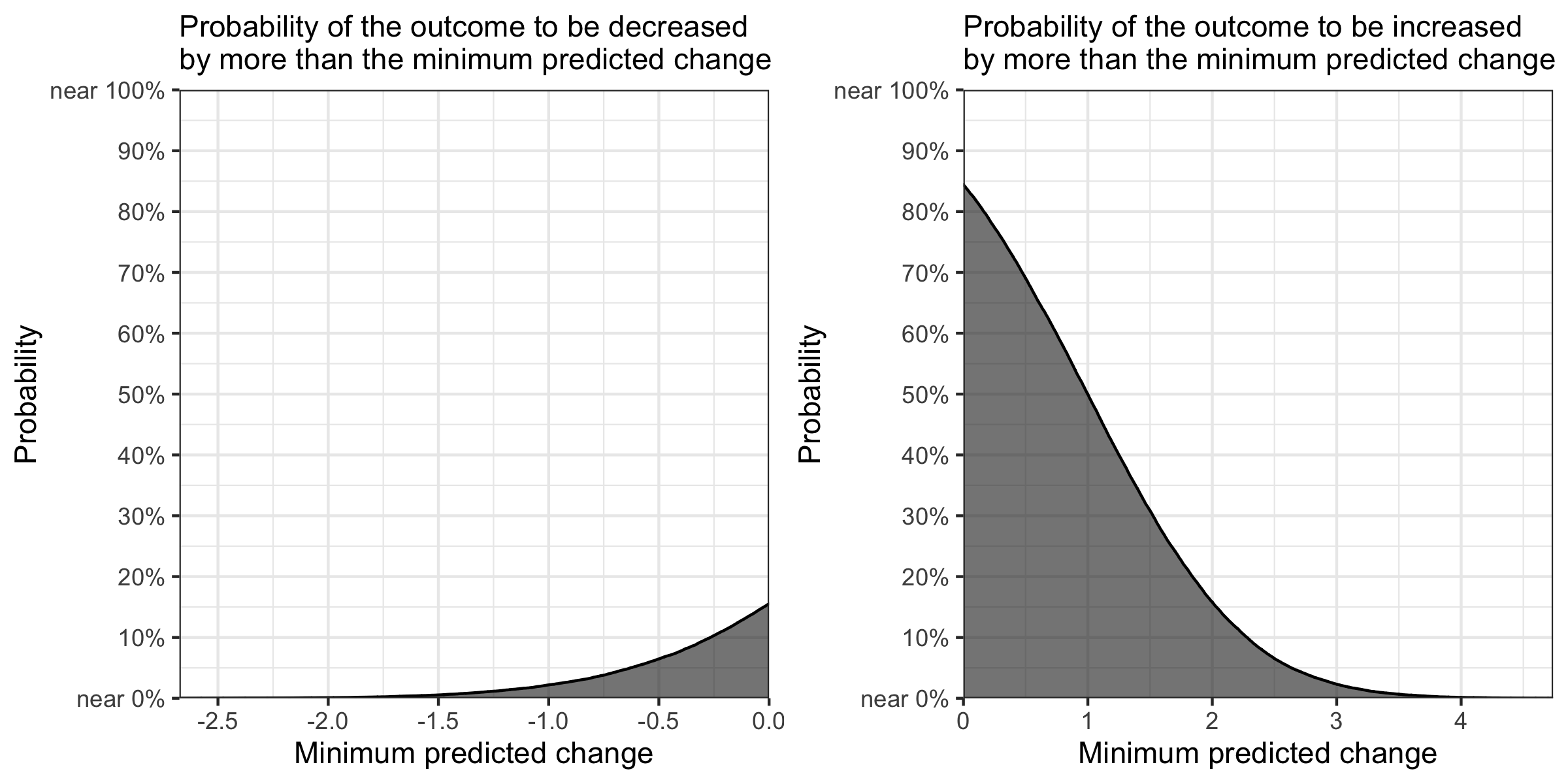}
  \centering
  \caption{Example of presenting a posterior as a complementary cumulative distribution plot.}
  \label{ccdfExa}
\end{figure}

Figure \ref{ccdfExa} is an example of the proposed approach. I use a distribution of 40,000 draws from a normal distribution with the mean of 1 and the standard deviation of 1. Let us assume these draws are those from a posterior distribution, or ``posterior samples,'' for the coefficient of a binary treatment variable in a linear regression, so that the values represent a change in the outcome variable. The x-axis is different minimum effect sizes, or different values of minimum predicted changes in the outcome; the y-axis is the probability of the effect being greater (in absolute terms) than a minimum predicted change. The y-axis has the adjective ``near'' on 0\% and 100\%, because the normal distribution is unbounded and, therefore, the plot of the posterior samples cannot represent the exact 0\% and 100\% probabilities.

From the figure, it is possible to see what the probability is of the effect being greater (in absolute terms) than a certain effect size. For example, we can say: ``The effect is expected to increase the outcome by greater than zero point with a probability of approximately 84\%.'' It is also clear that the positive effect is much more probable than the negative effect: 16\% probability for $\theta<0$ vs. 84\% probability for $\theta>0$. Finally, with a little more attention, we can also read the probability of a range of the effect size. For example, we can compute the probability that the effect increases the outcome by greater than one point and up to three points, as $P(\theta>1) - P(\theta>3) \approx .50 - .02 = .48$. This computation is useful if the effect sizes greater than a certain size are ``too large'' and counterproductive (e.g., the effect of a diet on weight loss).

\begin{figure}[t]
  \includegraphics[scale=0.15]{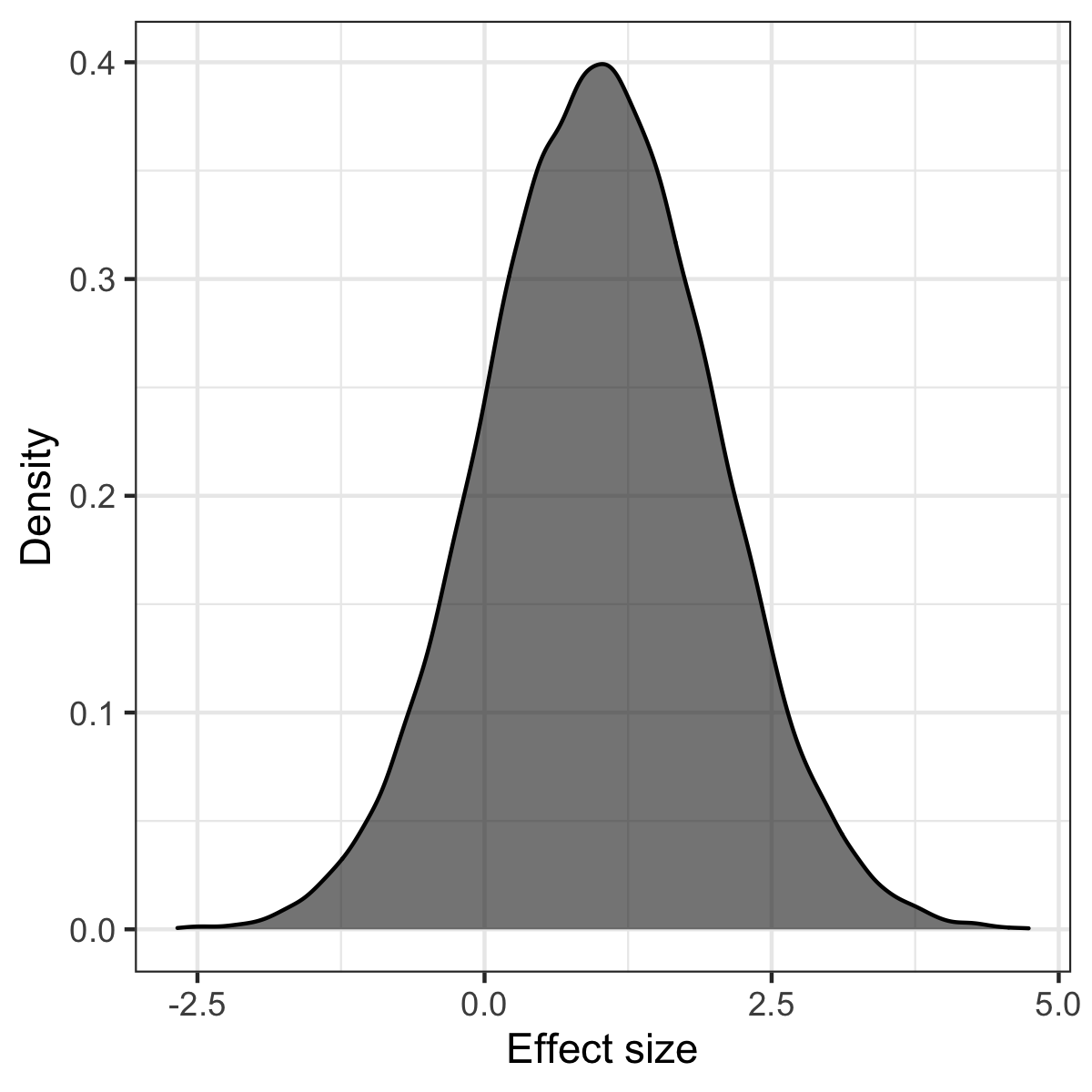}
  \centering
  \caption{Example of presenting a posterior as a probability density plot.}
  \label{pdfExa}
\end{figure}

For comparison, I also present the standard ways to interpret a posterior, using the same 40,000 draws as in Figure \ref{ccdfExa}: a probability density plot (Figure \ref{pdfExa}), a credible interval, and one-sided hypotheses (both in Table \ref{tableExa}). The probability density plot gives the overall impression that positive values are more probable than negative values. However, it is difficult to see with how much probability the causal factor has, for example, the positive effect greater than zero point. Unlike my approach, the y-axis is density rather than probability, and density is not an intuitive quantity.

\begin{table}[t]
\centering
\begin{tabular}{c c c c}
\hline
Mean & 95\% Credible Interval & $P(\theta>0)$ & $P(\theta<0)$\\
\hline
$1.00$ & [$-0.94$, $2.97$] & $0.84$ & $0.16$\\
\hline
\end{tabular}
\caption{Example of presenting a posterior as a credible interval or as one-sided hypotheses.}
\label{tableExa}
\end{table}

The credible interval is computed based on the conventional decision threshold of the 95\% credible level. It includes both negative and positive values and, therefore, the conventional approach leads to either of the two conclusions: that the effect is not statistically significant, or that the evidence is inconclusive \autocite{Gross2015, Kruschke2018a}.

The one-sided hypotheses are computed based on the decision threshold of the null effect, i.e., $P(\theta>0)$ and $P(\theta<0)$, as commonly done. Because of these decision thresholds, the information about the uncertainty is much more limited than what my approach presents in Figure \ref{ccdfExa}.

The use of decision thresholds also requires researchers to justify why these thresholds should be used, rather than, say, the 94\% credible interval or $P(\theta>0.1)$ and $P(\theta<-0.1)$. This problem does not apply to my approach. The proposed approach can be considered as a generalized way to use one-sided hypotheses. It graphically summarizes all one-sided hypotheses (up to the theoretical or practical limits), using different effect sizes as different thresholds based on each of which the probability of the effect is computed.

I note three caveats when the proposed approach is used. First, it takes up more space than presenting regression tables, the common presentation style in social science. Therefore, presenting all regression coefficients using my approach would be too cumbersome. Yet, the purpose of quantitative causal research in social science is usually to estimate the effect size of a causal factor or two, and the remaining regressors are controls to enable the identification of the causal effect(s) of interest. Indeed, it is often hard to interpret all regression coefficients causally because of complex causal mechanisms \autocite{Keele2020}. When researchers use matching instead of regression, they typically report only the effect of a treatment variable \autocite[1--2]{Keele2020}, although the identification strategy is the same -- selection on observables \autocite[321--22]{Keele2015}. Thus, if researchers are interested in estimating causal effects, they usually need to report only one or two causal factors. If so, it is not a problem even if my proposed approach takes more space than regression tables.

Second, if the effect size is scaled in a nonintuitive measure (such as a log odds ratio in logistic regression), researchers should convert it to an intuitive scale to facilitate interpretation \autocite[for details, see][]{Sarma2020}. For example, in the case of logistic regression, researchers can express an effect size as the difference in the predicted likelihood of an outcome variable between different treatment statuses.

Third, as all models are wrong, Bayesian models are also wrong; they are a simplification of reality. A posterior distribution is conditional on data used and model assumptions. It cannot be a reliable estimate, if the data used are inadequate for the purpose of research (e.g., a sample being unrepresentative of the target population or collected with measurement errors), and/or if one or more of the model assumptions are implausible (which is usually the case). Moreover, in practice a posterior usually needs to be computed by a Markov chain Monte Carlo (MCMC) method, and there is no guarantee that the resulting posterior samples precisely mirror the true posterior. Therefore, the estimated probability of an effect should not be considered as the ``perfect'' measure of uncertainty. For instance, even if the estimated probability of an effect being greater than zero were 100\% \textit{given the model and the computational method}, it should NOT be interpreted as the certainty of the effect \textit{in practice}.

Given these caveats (note that the second and third ones apply to any presentation formats), the same principle applies to my proposed approach as to the \textit{p}-value: ``No single index should substitute for scientific reasoning'' \autocite[132]{Wasserstein2016a}. What matters is not the ``trueness'' of a model but the ``usefulness'' of a model. The proposed approach makes a model more useful than the conventional approaches to evaluating and communicating the uncertainty of statistically estimated causal effects, in the following respects. First, it uses the probability of an effect as an intuitive quantity of uncertainty, for a better understanding and communication. Second, it does not require any decision thresholds for uncertainty measures or effect sizes, and allows researchers to play the role of information providers presenting the probabilities of different effect sizes as such.

\section{Application}
I exemplify the proposed approach by applying it to a previous social scientific study \autocite{Huff2016} and using its dataset \autocite{Huff2015a}. \textcite{Huff2016} collected a nationally representative sample of approximately 2,000 Polish adults and experimented on whether more violent methods of protest by an opposition group increase or decrease public support. Specifically, I focus on the two analyses in Figure 4 of \textcite{Huff2016}, which present (1) the effect of an opposition separatist group using bombing in comparison to occupation, and (2) the effect of an opposition separatist group using occupation in comparison to demonstrations, on the attitude of the experiment participants towards tax policy with respect to the fiscal autonomy of the region. The two treatment variables are measured dichotomously, while the attitude towards the tax policy as the dependent variable is measured on the percentage scale (0\%--100\%) asking how much regional income tax revenue should stay in the region after regional autonomy is settled. \textcite{Huff2016} use linear regression per treatment variable to estimate its average effect. The model is $Y=\beta_0+\beta_{1}D+\epsilon$, where $Y$ is the dependent variable, $D$ is the treatment variable, $\beta_{0}$ is the constant, $\beta_{1}$ captures the size of the average causal effect, and $\epsilon$ is the error term.

In the application, I convert the model to the following equivalent Bayesian linear regression model:

\begin{align*}
y_{i} & \sim Normal(\mu_{i}, \sigma),\\
\mu_{i} & = \beta_0 + \beta_1 d_{i},\\
\beta_{0} & \sim Normal(\mu_{\beta_0}=50,\sigma_{\beta_0}=20),\\
\beta_{1} & \sim Normal(\mu_{\beta_1}=0,\sigma_{\beta_1}=5),\\
\sigma & \sim Exponential(rate=0.5),
\end{align*}

\noindent
where $y_{i}$ and $d_{i}$ are respectively the outcome $Y$ and the treatment $D$ for an individual $i$; $Normal(\cdot)$ denotes a normal distribution; $\mu_{i}$ is the mean for $i$ and $\sigma$ is the standard deviation in the normal distribution likelihood. For the quantity of interest $\beta_{1}$, I use a weakly informative prior of $Normal(\mu_{\beta_1}=0,\sigma_{\beta_1}=5)$. This prior reflects the point that the original study presents no prior belief in favor of the negative or positive average effect of a more violent method by an opposition group on public support, as it hypothesizes both effects as plausible. For $\beta_{0}$, the constant term, I use a weakly informative prior of $Normal(\mu_{\beta_0}=50,\sigma_{\beta_0}=20)$. This prior means that, without the treatment, respondents are expected to have the median value of the dependent variable on average, but a large deviation from the median value is allowed in light of the data.\footnote{While the baseline conditions are different between the two analyses (occupation in the first analysis and demonstrations in the second analysis), the weakly informative nature makes the prior plausible for both analyses.} For $\sigma$ in the likelihood, I put a weakly informative prior of the exponential distribution $Exponential(rate=0.5)$, implying any stochastic factor is unlikely to change the predicted outcome value by greater than 10 points. I use four chains of MCMC process; per chain 20,000 iterations are done, the first 1,000 of which are discarded. The MCMC algorithm used is Stan \autocite{StanDevelopmentTeam2019b}, implemented via the rstanarm package version 2.21.1 \autocite{Goodrich2020a}. The $\hat{R}$ was approximately 1.00 for every estimated parameter, suggesting the models did not fail to converge. The effective sample size exceeded at least 30,000 for every estimated parameter.

Table \ref{tableHK} presents the results in a conventional way: the mean and 95\% credible interval of the posterior of $\beta_{1}$ for the model examining the effect of bombing in comparison to occupation, and those for the model examining the effect of occupation in comparison to demonstrations. While the mean is a negative value in both models, the 95\% credible interval includes not only negative values but also zero and positive values. This means the usual interval approach \autocite[e.g.,][]{Gross2015, Kruschke2018a} would lead us to conclude that the evidence is inconclusive.

\begin{table}[t]
\centering
\begin{tabular}{c c c c}
\hline
& Mean $\beta_{1}$ & 95\% Credible Interval & N\\
\hline
bombing vs. occupation & $-2.5$ & [$-5.5$, $0.5$] & 996\\
occupation vs. demonstration & $-0.5$ & [$-3.4$, $2.4$] & 985\\
\hline
\end{tabular}
\caption{Results using the 95\% credible interval. $\hat{R}\cong1.00$ for all parameters.}
\label{tableHK}
\end{table}

\begin{figure}[b]
  \includegraphics[scale=0.15]{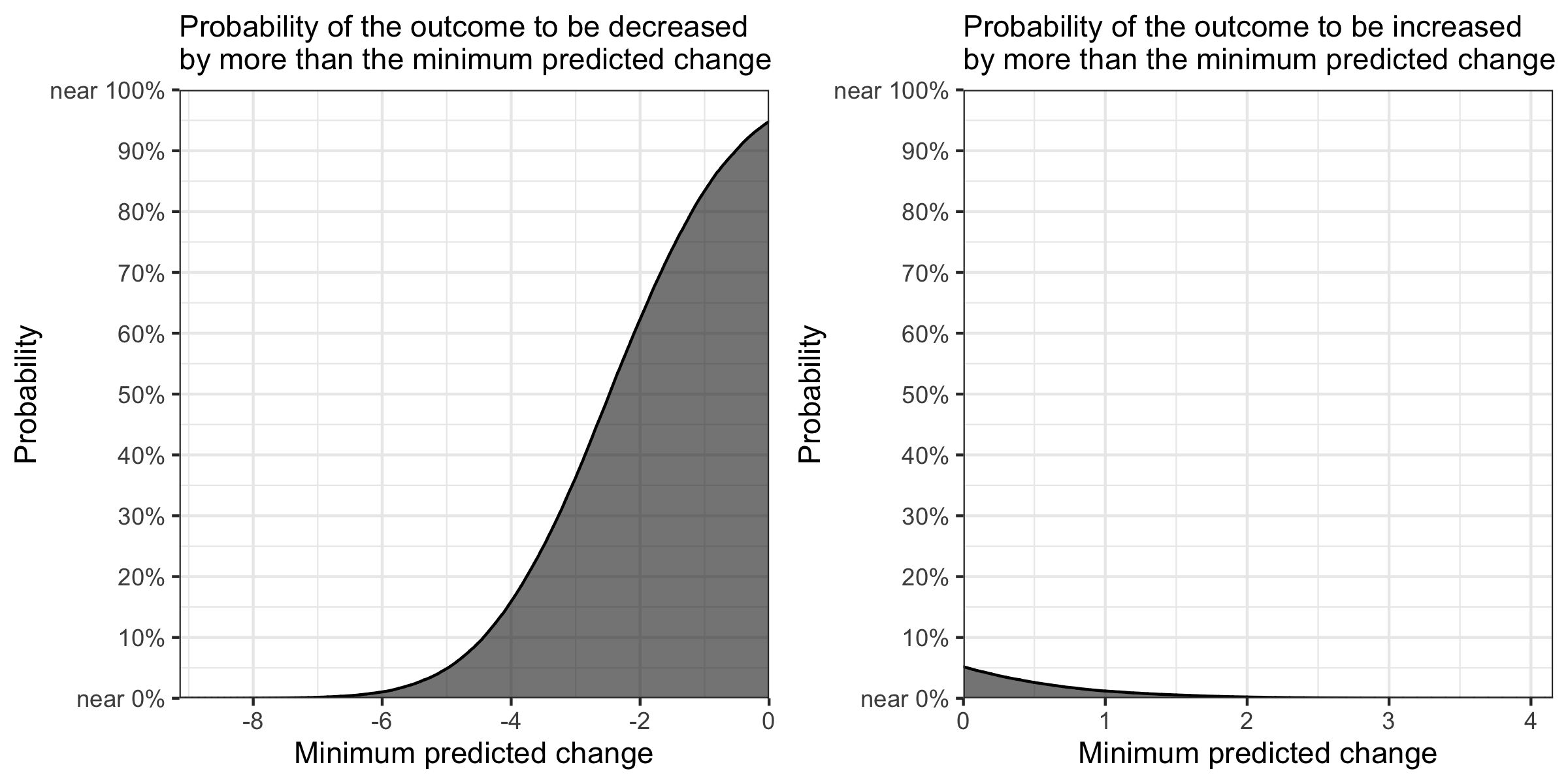}
  \centering
  \caption{Effect of bombing in comparison to occupation.}
  \label{ccdfBom}
\end{figure}

Figure \ref{ccdfBom} uses my proposed approach for the effect of bombing in comparison to occupation. It enables richer inference than the above conventional approach. If we focus, for example, on the probability of $\beta_{1}<0$, the model expects that if an opposition group uses bombing instead of occupation, there should be a greater than 0 point reduction in public support for the tax policy in favor of regional fiscal autonomy, with a probability of approximately 95\%. Thus, the negative effect of bombing is much more likely (95\% probability) than the positive effect (5\% probability).\footnote{The 95\% credible interval in Table \ref{tableHK} includes both positive and negative values because it is the two-tailed probability, while my proposed approach is based on one-tailed probabilities.}

The original conclusion was that the effect was not statistically significant at $p<5\%$, the threshold set at the pre-registration stage \autocite[1794--1795]{Huff2016}. However, the authors added a \textit{post hoc} caveat that if the threshold of statistical significance had been set at 10\%, the effect would have been regarded as statistically significant \autocite[1795]{Huff2016}. This interpretation is inconsistent both with the Fisherian paradigm of \textit{p}-values and with the Neyman-Person paradigm of \textit{p}-values \autocite{Lew2012}. According to the Fisherian paradigm, the preset threshold of statistical significance is unnecessary, because a \textit{p}-value in this paradigm is a local measure and not a global false positive rate -- the rate of false positives over repeated sampling from the same data distribution \autocite[1562--63]{Lew2012}. An exact \textit{p}-value should be interpreted as such, although it is difficult to interpret intuitively -- a p-value is the probability of obtaining data as extreme as, or more extreme than, those that are observed, given the null hypothesis being true \autocites[1560]{Lew2012}[131]{Wasserstein2016a}. According to the Neyman-Person paradigm, no \textit{post hoc} adjustment to the preset threshold of statistical significance should be made, because a \textit{p}-value in this paradigm is used as a global false positive rate and not as a local measure \autocite[1562--63]{Lew2012}. In my proposed approach, the estimates are more straightforward to interpret. We need no \textit{a priori} decision threshold of a posterior to determine significance; we can also \textit{post hoc} evaluate the probability of an effect \autocite[see][328]{Kruschke2015}.

\begin{figure}[t]
  \includegraphics[scale=0.15]{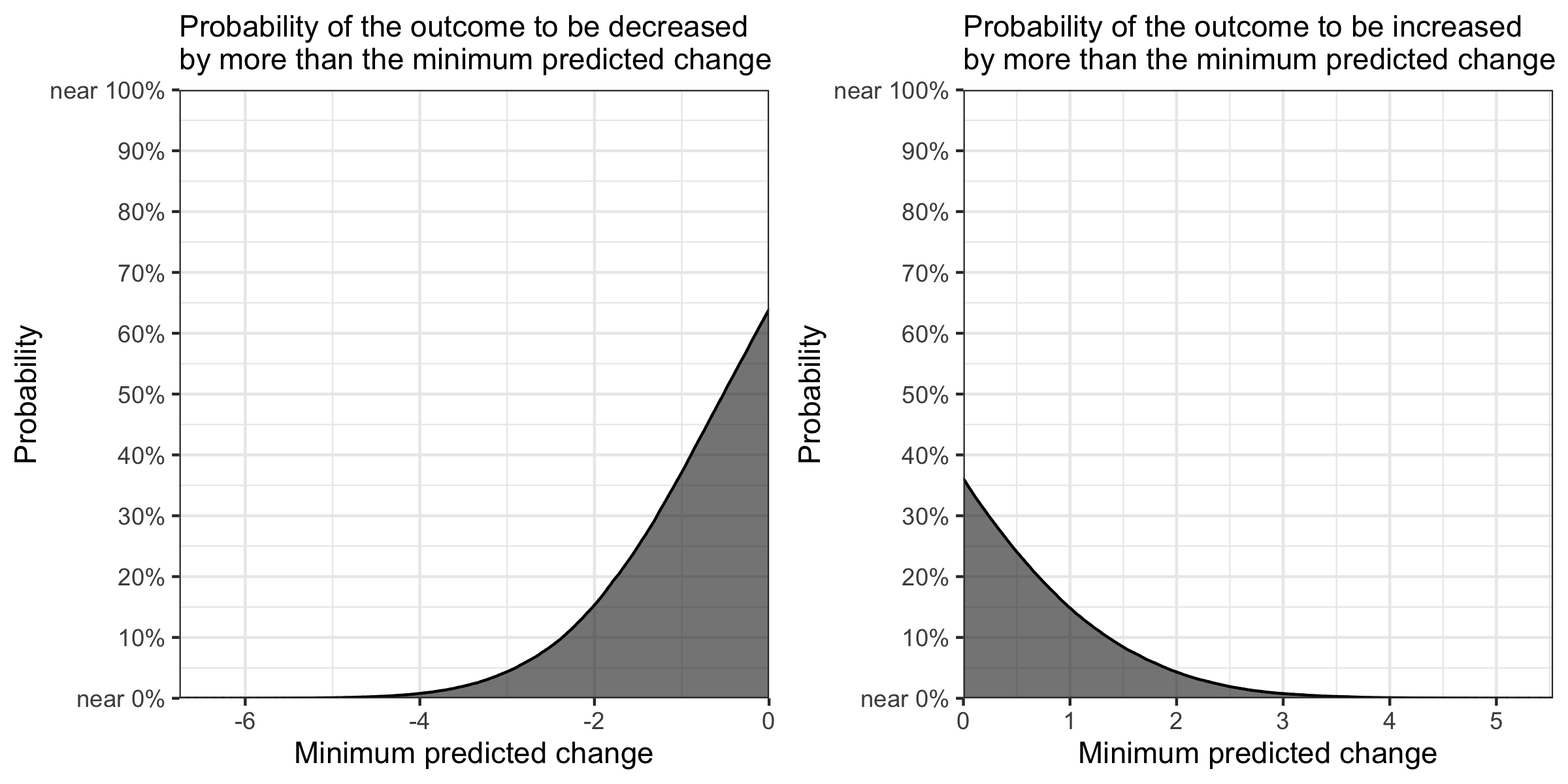}
  \centering
  \caption{Effect of occupation in comparison to demonstration.}
  \label{ccdfOcc}
\end{figure}

Figure \ref{ccdfOcc} illustrates my approach for the effect of occupation in comparison to demonstrations. The model expects that if an opposition group uses occupation instead of demonstrations, there should be a greater than 0 point reduction in public support for the tax policy in favor of regional fiscal autonomy, with a probability of approximately 64\%. In other words, the negative effect is more likely (64\% probability) than the positive effect (36\% probability), although the strength of evidence is lower than in the case of the comparison between bombing and occupation in Figure \ref{ccdfBom}. Meanwhile, the original conclusion was simply that the effect was not statistically significant at $p<5\%$ \autocite[1777, 1795]{Huff2016}.

In short, my approach presents richer information about the uncertainty of the statistically estimated causal effects, than the significance-vs.-insignificance approach taken by the original study. Thereby, it helps a better understanding and communication of the uncertainty of the average causal effects of protesting methods.

\section{Conclusion}
I have proposed an alternative approach for social scientists to present the uncertainty of statistically estimated causal effects: the probabilities of different effect sizes via the plot of a complementary cumulative distribution function. Unlike the conventional significance-vs.-insignificance approach and the standard ways to interpret a posterior distribution, the proposed approach does not require any decision threshold for the ``significance,'' ``confidence,'' or ``credible'' level of uncertainty or for the effect size beyond which the effect is considered practically relevant. It therefore allows researchers to be agnostic about these decision thresholds and justifications for them. Researchers play the role of information providers and present different effect sizes and their associated probabilities as such. Through the application to a previous study, I have shown that the proposed approach presents richer information about the uncertainty of statistically estimated causal effects than the conventional significance-vs.-insignificance approach, enabling a better understanding and communication of the uncertainty.

\section*{Acknowledgments}
I would like to thank Johan A. Dornschneider-Elkink, Jeff Gill, Zbigniew Truchlewski, Alexandru Moise, and participants in the 2019 PSA Political Methodology Group Annual Conference, the 2019 EPSA Annual Conference, and seminars at Dublin City University and University College Dublin, for their helpful comments. I would like to acknowledge the receipt of funding from the Irish Research Council (the grant number: GOIPD/2018/328) for the development of this work. The views expressed are my own unless otherwise stated, and do not necessarily represent those of the institutes/organizations to which I am/have been related.

\section*{Supplemental Materials}
The R code to reproduce the results in this article, and the R package to implement the proposed approach (``ccdfpost''), are available on my website at \url{https://akisatosuzuki.github.io}.

\printbibliography

\end{document}